\documentclass{article}

\usepackage{PRIMEarxiv}

\usepackage[utf8]{inputenc} 
\usepackage[T1]{fontenc}    
\usepackage{hyperref}       
\usepackage{url}            
\usepackage{booktabs}       
\usepackage{amsfonts}       
\usepackage{nicefrac}       
\usepackage{microtype}      
\usepackage{lipsum}
\usepackage{fancyhdr}       
\usepackage{graphicx}       
\graphicspath{{media/}}     
\usepackage{subcaption}
\usepackage{amsmath,amssymb,amsfonts}
\usepackage{algorithm}
\usepackage{algpseudocode}
\usepackage{tabularx, makecell, multirow, diagbox} 
\usepackage[acronym]{glossaries}
\usepackage{xspace}

\glsdisablehyper
\newacronym{pdr}{PDR}{Packet Delivery Ratio}
\newacronym{rss}{RSS}{Received Signal Strength}
\newacronym{snr}{SNR}{Signal-to-Noise Ratio}
\newacronym{ee}{EE}{Energy Efficiency}
\newacronym{adr}{ADR}{Adaptive Data Rate}
\newacronym{rl}{RL}{Reinforcement Learning}
\newacronym{drl}{DRL}{Deep Reinforcement Learning}

\title{FAST-LoRa: An Efficient Simulation Framework for Evaluating LoRaWAN Networks and Transmission Parameter Strategies
}

\author{
Laura Acosta García\thanks{Telematics Engineering Group, Technical University of Cartagena, 30202 Spain.} \and
Juan Aznar Poveda\thanks{Distributed and Parallel Systems Group, University of Innsbruck, 6020 Austria.} \and
Fabian Margreiter\footnotemark[2] \and
Antonio-Javier García Sánchez\footnotemark[1] \and
Joan García Haro\footnotemark[1] \and
Thomas Fahringer\footnotemark[2] \and
José Lorente López\thanks{Mobile Communications Systems Group, Technical University of Cartagena, 30202 Spain.} \and
José-Víctor Rodríguez\footnotemark[3]
}
\newcommand{\system}{FAST-LoRa\xspace}

\begin{document}

\maketitle

\begin{abstract}
The Internet of Things (IoT) has transformed many industries, and LoRaWAN (Long Range Wide Area Network), built on LoRa (Long Range) technology, has become a crucial solution for enabling scalable, low-cost, and energy-efficient communication in wide-area networks. Simulation tools are essential for optimizing the transmission parameters and, therefore, the energy efficiency and performance of LoRaWAN networks. While existing simulation frameworks accurately replicate real-world scenarios by including multiple layers of communication protocols, they often imply significant computational overhead and simulation times. To address this issue, this paper introduces \system, a novel simulation framework designed to enable fast and efficient evaluation of LoRaWAN networks and selection of transmission parameters. \system streamlines computation by relying on analytical models without complex packet-level simulations and implementing gateway reception using efficient matrix operations. Rather than aiming to replace discrete-event simulators, \system is intended as a lightweight and accurate approximation tool for evaluating transmission parameter strategies in scenarios with stable traffic patterns and uplink-focused communications. In our evaluation, we compare \system with a well-established simulator using multiple network configurations with varying numbers of end devices and gateways. The results show that \system achieves similar accuracy in estimating key network metrics, even in complex scenarios with interference and multi-gateway reception, with a Mean Absolute Error (MAE) of 0.940 $\times 10^{-2}$ for the Packet Delivery Ratio (PDR) and 0.040 bits/mJ for Energy Efficiency (EE), while significantly reducing computational time by up to three orders of magnitude.
\end{abstract}

\maketitle

\section{Introduction}
\label{sect:intro}
The Internet of Things (IoT) has transformed how devices communicate, enabling advancements in sectors such as healthcare, agriculture, and smart cities. One of the technologies driving this change is LoRa (Long Range), a low-power wide-area network (LPWAN) that allows for long-range communication with minimal energy consumption  \cite{gkotsiopoulos2021performance, sundaram2019survey}. LoRaWAN, the protocol built on LoRa, further enhances this capability, making it well-suited for battery-powered IoT devices. Due to its scalability and cost-effectiveness, LoRaWAN is becoming an essential part of modern IoT systems, supporting a wide range of applications from environmental monitoring to industrial automation.

An essential mechanism for optimizing LoRaWAN network performance is the Adaptive Data Rate (ADR). The ADR adjusts the transmission parameters of devices based on network conditions, such as signal strength, interference, and network load. By dynamically modifying transmission power and the spreading factor, the ADR ensures that devices maintain efficient communication while balancing critical factors such as range, data rate, and energy consumption. As LoRaWAN networks expand and become more complex, optimizing the ADR becomes more challenging. However, it remains crucial for maintaining appropriate network performance and energy consumption \cite{benkahla2019enhanced, farhad2020enhanced, moysiadis2021extending, anwar2021rm, jouhari2023deep, muthanna2022deep, mhatre2022dynamic, yazid2022reinforcement, acosta2023dynamic}. Optimizing LoRaWAN networks in real-world environments often requires significant financial investments in hardware, such as gateways and end devices, and infrastructure, including installation and maintenance \cite{sotiriadis2014towards}. Additionally, the inherent variability of environmental conditions, interference, and the spatial distribution of nodes complicates a comprehensive evaluation of network performance.

In this context, simulation tools offer a cost-effective solution for testing and optimizing LoRaWAN network configurations without the need for extensive real-world deployments. These tools help evaluate various network configurations and data transmission strategies. Several simulation tools have been developed to support LoRaWAN networks \cite{bor2016lorasim, bor2016lora, slabicki2022flora, slabicki2018adaptive, magrin2016network, magrin2016ns3, UniCTARS44, ta2019lora, serati2022adr, marais2019review}, providing features that simulate network topology, protocol layers, signal propagation, interference, mobility, traffic management, Quality of Service (QoS), security, and energy consumption. However, these tools often imply significant computational overhead, particularly in large-scale or dynamic environments, which limits their capacity to support rapid, iterative optimizations. This issue becomes particularly important when attempting to dynamically fine-tune ADR settings efficiently in complex, large, and dense networks.

To address this limitation, we introduce \system, a novel simulation framework designed to enable the efficient optimization of transmission parameters in LoRaWAN networks. \system streamlines the simulation process by optimizing core transmission parameters, such as the transmission rate, power, and spreading factor. This approach permits rapid and iterative testing of ADR settings and network configurations without the computational burden associated with simulating complex network behaviors that are not directly relevant to transmission optimization. To achieve this efficiency, \system simplifies the simulation process by assuming a single-channel configuration, abstracting packet-level behavior using analytical models, and computing gateway receptions through matrix-based operations. While \system is not designed to replace detailed packet-level discrete-event simulators, it provides a lightweight and accurate alternative for evaluating transmission strategies in scenarios where traffic patterns are stable and communication is predominantly uplink-focused. \system enables the simulation of large-scale LoRaWAN networks with numerous devices and gateways, making it ideal for fast evaluations of transmission parameter selection strategies and network layouts, e.g., planning of gateway placement. The contributions of this paper are summarized as follows:

\begin{itemize}
    \item We present \system, a novel open-source simulation framework for efficient LoRaWAN network analysis and transmission parameters selection.
    \item The proposed framework enables the efficient evaluation of large-scale LoRa networks without packet-level simulation overhead.
    \item \system efficiently estimates key performance metrics, such as the Packet Delivery Ratio (PDR) and Energy Efficiency (EE), streamlining transmission parameter optimization.
    \item We evaluate \system across various network configurations and validate it with the well-established simulator, FLoRa. Our results demonstrate that \system achieves comparable accuracy in the Packet Delivery Ratio (PDR) and energy efficiency (EE), with a Mean Absolute Error (MAE) of 0.940 $\times 10^{-2}$ for the PDR and 0.040 bits/mJ for EE, while drastically reducing execution time by up to three orders of magnitude in dense network scenarios. These results are obtained in scenarios aligned with the modeling assumptions of \system, such as fixed traffic patterns and uplink-dominated communication.
\end{itemize}

The rest of this paper is organized as follows. Section~\ref{sect:rw} reviews and discusses related work. Section~\ref{sect:model} describes the analytical model of \system, including the network description, transmission time calculation, criteria for determining successful transmissions and receptions, and implementation details. Section \ref{sec:setup} describes the evaluation setup, metrics, and baselines used. In Section~\ref{sec:results}, we validate the correctness of our simulation framework by comparing it to a well-established framework. Finally, Section~\ref{sect:conc_fw} summarizes the major findings of this study and outlines future work.

\section{Related Work}
\label{sect:rw}
In this section, we review the state of the art in selecting transmission parameters in LoRaWAN networks and how this is achieved by commonly used LoRaWAN simulators. 

\subsection{Transmission parameter selection}
Numerous research efforts have been made to select transmission parameters in LoRaWAN networks \cite{benkahla2019enhanced, farhad2020enhanced, moysiadis2021extending, anwar2021rm, jouhari2023deep, muthanna2022deep, mhatre2022dynamic, yazid2022reinforcement, acosta2023dynamic, acosta2024proactive, acosta2025adrl}. For example, the authors of \cite{anwar2021rm} introduced RM-ADR, a novel mechanism that operates on the network server and end devices. The end device algorithm reduces retransmission attempts, while the network server leverages previous frame reception power to dynamically select the spreading factor ($f$) and transmission parameters. Other techniques, such as Gaussian-based filtering (G-ADR) and the Exponential Moving Average (EMA-ADR), have been explored to smooth out rapid Signal-to-Noise Ratio (SNR) fluctuations \cite{farhad2020enhanced, farhad2022hadr}. Many of these works leverage artificial intelligence to dynamically learn new transmission policies over time. For instance, \cite{yazid2022reinforcement} employs reinforcement learning to select energy efficiency and ensure reliable packet delivery. The work in \cite{muthanna2022deep} proposes an intelligent routing policy to find suitable transmission parameters and move beyond the traditional star topology to improve network performance. In \cite{acosta2023dynamic}, the authors employ Deep Reinforcement Learning (DRL) to determine the optimal spreading factor and transmission power combination, reducing energy consumption while maintaining an adaptive Bit Error Rate (BER) threshold. Similarly, \cite{mhatre2022dynamic, acosta2025adrl} uses DRL to improve energy efficiency by dynamically adjusting transmission settings. However, the vast majority of these works either propose their custom simulator \cite{acosta2023dynamic, acosta2024proactive, acosta2025adrl, jouhari2023deep, yazid2022reinforcement}, which often lacks rigorous validation, or employ well-known simulators, which often imply notable computational overhead and hamper the selection process.

\subsection{State-of-the-art LoRaWAN simulators}
Simulation frameworks are frequently employed to find suitable transmission parameters and energy consumption in LoRa networks. One of the most widely used simulation tools for LoRa networks is LoRaSim \cite{bor2016lorasim, bor2016lora}, a simulation tool that focuses on key aspects such as collisions, spreading factors, and gateway coverage. Recent improvements to LoRaSim, such as those proposed in \cite{francisco2021improving}, address some of its limitations by incorporating more realistic propagation models and the capture-effect phenomenon. These enhancements help improve the accuracy of network behavior simulations, particularly in dense environments. LoRaWANSIM \cite{pop2017does}, an extension of LoRaSim, adds features such as bidirectional communication and medium access control (MAC) layer improvements. However, although LoRaSim and LoRaWANSIM are effective for evaluating ADR mechanisms, their focus on the physical and MAC layers restricts their ability to fully capture application-layer dynamics. FLoRa \cite{slabicki2022flora, slabicki2018adaptive}, built on the OMNeT++ simulation framework \cite{opensim2019omnetpp} and the INET framework \cite{opensim2023inet}, provides a more comprehensive simulation environment, offering detailed models for communication delays, energy consumption, and network dynamics. Its modular architecture allows for the simulation of multiple layers of the LoRaWAN protocol stack—including the physical, MAC, network, and application layers—enabling cross-layer analysis and the integration of custom ADR mechanisms. Similarly, the ns-3 simulator \cite{nsnam2024ns3}, an open-source platform that supports a wide range of wireless communication technologies, including LoRaWAN, offers extensive flexibility for simulating complex, multi-technology scenarios \cite{magrin2016network, magrin2016ns3}. Its ability to model physical and higher-layer interactions makes it a powerful tool for evaluating \gls{adr} mechanisms in integrated environments. However, FLoRa and ns-3 face similar limitations due to the significant computational resources required for simulating individual transmissions. This makes them impractical for tasks requiring rapid iterations, such as reinforcement learning training \cite{acosta2023dynamic, jouhari2023deep, muthanna2022deep}, or for evaluating critical network performance metrics like energy efficiency, communication reliability, and the packet delivery rate. 

LWN-Simulator \cite{UniCTARS44} configures LoRaWAN in urban environments, modeling end device mobility, interference, and gateway placement to evaluate ADR mechanisms. Similarly, LoRa-MAB \cite{ta2019lora} employs Multi-Armed Bandit (MAB) algorithms for decentralized resource allocation, enabling end devices to make autonomous decisions. In addition, \cite{serati2022adr} proposes a link-based ADR approach that enhances the PDR in mobile scenarios, overcoming the traditional ADR. However, these simulators often demand high computational resources and face scalability issues in large or highly dynamic networks. Other tools focus on specific aspects: the MATLAB-based simulator \cite{centenaro2017impact} assesses confirmed traffic impact but is closed-source; LoRaEnergySim \cite{callebaut2019cross} models energy consumption, supporting ADR and two-channel models; and a LoRa/Sigfox simulator \cite{maartenw68} analyzes physical layer characteristics. Despite their contributions, these simulators prioritize specific network aspects rather than comprehensive network optimization.

\section{\system simulation framework}
\label{sect:model}
\system is a simulation framework designed to provide fast LoRaWAN network evaluations and find suitable transmission parameters efficiently. Unlike traditional simulators that require high computational power to consider numerous aspects and multiple layers of communication protocols, \system adopts a statistical approach \cite{eigner2021interference, zhang2023multiagent} that efficiently obtains network performance metrics, such as the Packet Delivery Ratio (PDR) and Energy Efficiency (EE). By exclusively focusing on essential aspects related to transmission parameter selection, our approach enables faster simulations, making it particularly useful for online optimization techniques or machine learning methods that require a large number of iterations. This efficiency is due to modeling simplifications, such as focusing only on transmission-related metrics, replacing packet-level interactions with analytical models, and assuming a single-channel configuration. As a result, \system is best suited for scenarios with fixed traffic patterns and uplink-dominated communication, where such simplifications still provide accurate approximations of network behavior.

\subsection{Network model}
\begin{figure}
    \centering
    \includegraphics[width=0.75\linewidth]{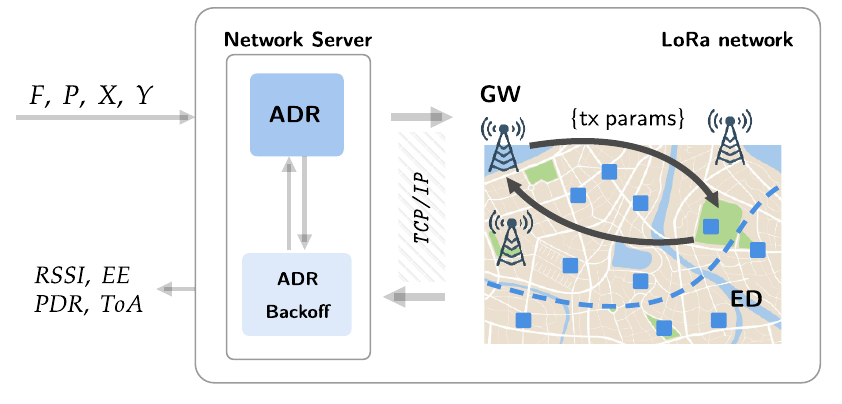}
    \caption{Overview of \system, including the network description, ADR and backoff mechanisms, input parameters, and simulation results.}
    \label{fig:overall_diagram}
\end{figure}

\system leverages a realistic analytical network model to efficiently estimate key performance metrics. We consider a LoRa network consisting of $N$ end devices (EDs) and $K$ gateways (GWs). Each ED $i$ is assigned a transmission power $p_i \in \mathcal{P}$, a spreading factor $f_i \in \mathcal{F}$, and a coding rate (CR) $r_i \in \mathcal{R}$. Regarding the network topology, we consider the star topology typically used in LoRaWAN networks, where all the EDs communicate with the gateways in coverage, which are in turn interconnected through a central  Network Server (NS). The NS controls and optimizes the communication between the end devices and the gateways. Although LoRa networks support the simultaneous use of multiple channels, cross-channel interference is generally negligible when the frequency ranges of the channels do not overlap \cite{yu2024resolve}. This condition holds for most LoRaWAN channel plans defined by the LoRa Alliance \cite{loraalliance2022regionalparameters}. Consequently, \system focuses on simulating only intra-channel interference, assuming that all devices operate on the same channel. This simplification results in more efficient simulation while still providing a realistic representation of the network's performance. To better illustrate how the network components and the various mechanisms interact within \system, an overview is presented in Figure~\ref{fig:overall_diagram}.

\subsection{Time-on-air model}
Time-on-air ($T_i$) is a key metric for evaluating performance, particularly concerning interference and energy efficiency. Time-on-air refers to the duration that a wireless transmission occupies the communication medium, from the start of the preamble until the end of the payload transmission. The time-on-air required by ED $i$ on a channel with bandwidth $B$ using a spreading factor of $f_i \in \mathcal{F}$ and CR $r_i \in \mathcal{R}$ for transmitting a single LoRa packet with $L$~bytes of payload can, according to \cite{semtech2020sxdatasheet}, be calculated as:
 
\begin{equation} \label{eq:packet-transmission-time}
    T_i = T^{pr}_{i} + T^{pl}_{i}
    \end{equation}

where $T_i$ is the total time-on-air for ED $i$, consisting of preamble time ($T^{pr}_{i}$) and payload transmission time ($T^{pl}_{i}$). The time to transmit a single symbol ($T^{sym}_{i}$) can be calculated as \cite{semtech2020sxdatasheet}:
 
\begin{equation} \label{eq:symbol-transmission-time}
    T^{sym}_{i} = \frac{2^{f_i}}{B}
\end{equation}

where $f_i$ is the spreading factor of ED $i$, and $B$ is the bandwidth of the channel. The time to transmit the LoRa preamble ($T^{pr}_{i}$), where the number of preamble symbols $n_{pr}$ is a fixed value, can be calculated as follows \cite{semtech2020sxdatasheet}:
 
\begin{equation} \label{eq:preamble-transmission-time}
    T^{pr}_{i} = (n_{pr} + 4.25) \, T^{sym}_{i}
\end{equation}

The time to transmit the payload ($T^{pl}_{i}$) is calculated as \cite{semtech2020sxdatasheet}:
 
\begin{equation} \label{eq:payload-transmission-time}
    T^{pl}_{i} = n_{pl} \, T^{sym}_{i}
\end{equation}

 where $n_{pl}$ is the number of symbols needed to encode the payload information and is calculated as \cite{semtech2020sxdatasheet}:
 
\begin{equation} 
\label{eq:payload-symbols}
        n_{pl} = 8 + \text{max}\left(\left\lceil\frac{8L - 4f_{i} + 28 + 16CRC - 20H}{4(f_{i} - 2DE)}\right\rceil(r_i+4), 0\right)
\end{equation}
 
where $L$ is the payload size in bytes, $CRC$ is 1 if the packet contains a checksum and 0 otherwise, $H$ is 0 if the packet header is enabled and 1 otherwise, and $DE$ is 1 if low data rate optimization is enabled and 0 otherwise. In our approach, the time-on-air calculations are specifically focused on uplink messages where we model the transmission of packets from EDs to GWs. Downlink messages are not explicitly simulated under the assumption that their payloads are small and the corresponding energy consumption in reception mode is negligible \cite{acosta2025adrl}. This simplification is typically sufficient to evaluate the performance of different ADR algorithms, which primarily rely on uplink behavior \cite{bor2016lora, haxhibeqiri2018survey, LoRaSim, magrin2017performance, zhang2023multiagent}. We also assume that the decoding probability of downlink messages is similar to their predecessor uplink messages \cite{coutaud2021lora}.

\subsection{Transmission reliability model}
In a LoRa network, transmission reliability can be determined by two key factors: (i) the \gls{rss} at the gateway must exceed its receiver sensitivity, and (ii) the packet must not be corrupted by interference from other end device transmissions \cite{gao2019towards}. The PDR represents the probability that a packet from ED $i$ is successfully received by GW $k$, and it can be calculated as the product of two probabilities: that the \gls{rss} is sufficient for successful reception ($\psi_{ik}$) and that the packet is not corrupted by interference ($\zeta_{ik}$)  \cite{zhang2023multiagent}:
 
\begin{equation} 
\label{eq:pdr-base}
    PDR_{ik} = \psi_{ik} \cdot \zeta_{ik}
\end{equation}

To calculate the probability that the RSS is sufficient for reception ($\psi_{ik}$), we express the RSS as the difference between a deterministic component $z_{ik}$ and a random noise component $N_{ik}$, where $RSS_{ik} = z_{ik} - N_{ik}$, with $z_{ik} = p_i - \overline{PL}(d_0) - 10 \gamma \log_{10} \left( \frac{d_{ik}}{d_0} \right)$ and $N_{ik} \sim \mathcal{N}(0, \sigma)$, a Gaussian random variable, representing the noise. In this paper, we consider the log-distance path loss model since it is suitable for rural and urban environments. It considers the transmission power, distance between the end device and the gateway, and effects of signal attenuation due to fading. Using the log-distance model, the \gls{rss} of a transmission can be calculated as follows:
 
\begin{equation} 
\label{eq:rss-1}
    RSS_{ik} = p_i - \overline{PL}(d_{0}) - 10 \gamma \log_{10}(\frac{d_{ik}}{d_0}) - N_{\sigma}
\end{equation}
 
where $p_i \in \mathcal{P}$ is the transmission power setting of ED $i$; $\overline{PL}(d_{0})$ is the mean path loss at a reference distance $d_0$, usually selected under line-of-sight (LoS) conditions; $\gamma$ is the path loss exponent, indicating how quickly the signal attenuates in a particular environment; $d_{ik}$ is the euclidean distance between ED $i$ and GW $k$; and $N_{\sigma}$ is a Gaussian random variable with a zero mean and standard deviation of $\sigma$, reflecting the signal variation caused by shadow fading. The probability that the RSS exceeds receiver sensitivity $\eta_{f_i}$ for a given spreading factor $f_i$ is then calculated using the error function (erf) as follows:

\begin{equation} 
    \begin{gathered}
    \label{eq:prob-signal-strength}
    \psi_{ik} = \mathbb{P}(RSS_{ik} \geq \eta_{f_{i}})\\ = \mathbb{P}(N_{ik} \leq z_{ik} - \eta_{f_{i}}) =
    \frac{1}{2} + \frac{1}{2} \, \text{erf}(\frac{z_{ik}-\eta_{f_i}}{\sqrt{2}\sigma})
    \end{gathered}
\end{equation}

where $\eta_{f_{i}}$ describes the sensitivity of a receiving GW at spreading factor $f_{i}$, and $\text{erf}$ denotes the Gauss error function.

\subsection{Interference model}
\begin{figure}[ht!]
    \centering
    \includegraphics[width=0.75\linewidth]{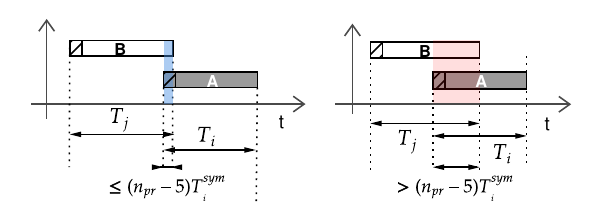}
    \caption{Illustration of overlap and interference between two transmissions A and B.}
    \label{fig:transmission-interference}
\end{figure}

To accurately model the interference across transmissions, we first calculate the likelihood of overlapping transmissions. Given two EDs, $i$ and $j$, the probability that ED $j$ will interfere with ED $i$ is based on the time overlap of their transmissions. We assume that the transmission from an ED can be successfully recognized if at least the last five preamble symbols are received without interference \cite{bor2016lora}, as shown in Figure~\ref{fig:transmission-interference}.

Therefore, the time interval during which ED $j$ must not transmit to avoid interfering with a transmission from ED $i$ can be calculated as:

\begin{equation} 
\label{eq:interference-interval}
    T^{'}_{ij} = T_j + T_i - (n_{pr} - 5)T^{sym}_{i}
\end{equation}

where $T_j$ and $T_i$ are the transmission durations of EDs $j$ and $i$, respectively. With this, we calculate the probability that ED $j$ interferes with ED $i$, considering the exponential nature of packet generation in the network \cite{slabicki2022flora, zhang2023multiagent}:

\begin{equation} 
\label{eq:interference-probability-single}
    h_{ij} = 1 - e^{-\lambda T^{'}_{ij}}
\end{equation}
 
where $\lambda$ is the mean packet generation rate.

When calculating the total interference probability, we also need to take into account the modulation characteristics of LoRa. The robustness of LoRa modulation allows for successful decoding even if the RSS is below the noise floor, provided that the signal-to-interference ratio (SIR) is above a certain threshold, which depends on the spreading factor employed by transmissions from the affected ED $i$ and the interfering ED $j$ \cite{amichi2020joint, toro2021modeling, xu2022x}. Then, the probability that the transmission from ED $i$ will be corrupted by interference from ED $j$ is calculated as:
 
\begin{equation}
    \begin{gathered}
        \mathbb{P}(RSS_{ik} - RSS_{jk} < \omega_{f_{i}f_{j}})\\ =
        \mathbb{P}(N_{jk} - N_{ik} < \omega_{f_{i}f_{j}} - (z_{ik} - z_{jk})) \\=
        \frac{1}{2} + \frac{1}{2} \, \text{erf}\left(\frac{\omega_{f_{i}f_{j}} - (z_{ik} - z_{jk})}{2\sqrt{2}\sigma}\right) 
    \end{gathered}
\end{equation}
 
where $\omega_{f_{i}f_{j}}$ is the signal-to-interference ratio threshold, and $\text{erf}$ is the Gauss error function, which can be applied since both $N_{jk}$ and $N_{ik}$ are Gaussian random variables $\mathcal{N}(0, \sigma)$, combinable with another Gaussian random variable $\mathcal{N}(0, 2\sigma)$.

Considering all the end devices in a network, the probability that a packet from ED $i$ received at GW $k$ will not be corrupted by transmissions from other end devices can be calculated as:
 
\begin{equation} 
    \label{eq:interference-probability}
    \zeta_{ik} = \prod_{j \in \{1..N\}, i \ne j}(1 - h_{ij} \, \mathbb{P}(RSS_{ik} - RSS_{jk} < \omega_{f_{i}f_{j}}))
\end{equation}

\subsection{Performance metrics}
Finally, the overall PDR of ED $i$ can be calculated by considering all the available gateways GW, and combining the individual PDRs for each GW. As defined in Eq.~\eqref{eq:pdr-base}, the PDR between ED $i$ and each GW $k$ is given by the product of the probabilities that the RSS will be sufficient for reception ($\psi_{ik}$) and that the packet will not be corrupted by interference ($\zeta_{ik}$). Once these components are calculated for each pair of EDs and GWs, the overall PDR for ED $i$ is obtained by combining the individual PDRs from each gateway using the following equation:

\begin{equation} 
    \label{eq:pdr-overall}
    PDR_{i} = 1 - \prod_{k \in \{1..K\}}(1 - PDR_{ik})
\end{equation}

This equation aggregates the individual PDRs from each gateway, providing the overall probability that a packet from ED $i$ will be successfully received by at least one GW in the network, considering the interference and signal quality at each GW. It is possible to sample the \gls{rss} of an individual packet transmission from end device $i$ arriving at gateway $k$ as:
\nopagebreak
\begin{equation} 
    \label{eq:rss-sampled}
    \hat{RSS_{ik}} =
        \begin{aligned}
        \begin{cases}
            RSS_{ik} &\text{ if } RSS_{ik} \geq \eta_{f_i} \text{, with probability } \zeta_{ik}\\
            -\infty &\text{ otherwise}
        \end{cases}
        \end{aligned}
\end{equation}
\nopagebreak
where $\hat{RSS_{ik}}$ represents the sampled \gls{rss}, which is calculated as the \gls{rss} (including random signal attenuation by flat fading) if it exceeds receiver sensitivity $\eta_{f_i}$ for the used spreading factor and also considering possible corruption by interference with transmissions from other end devices $\zeta_{ik}$. An \gls{rss} of $-\infty$ indicates that the packet has not been decoded successfully in the chosen sample.

Similarly, the \gls{snr} and Energy Efficiency (EE) are two crucial metrics that impact transmission performance. The SNR determines the quality of the received signal at the GW, which influences the likelihood of successful demodulation, while EE reflects the efficiency of energy consumed during data transmission. The \gls{snr} of a transmission from ED $i$ arriving at GW $k$ can generally be calculated as:
 
\begin{equation} \label{eq:snr-base}
    SNR_{ik} = RSS_{ik} - P_{noise}
\end{equation}

where $P_{noise}$ represents the noise floor in dBm. LoRa's modulation allows for successful demodulation even when the \gls{rss} is below the noise floor, depending on the spreading factor used \cite{farhad2022hadr, farhad2020enhanced, moysiadis2021extending, slabicki2018adaptive}. The minimum SNR required for successful demodulation for each spreading factor ($f$) is listed in Table~\ref{tab:min-snr}.
 
\begin{table}[ht!]
    \centering
    \setlength{\tabcolsep}{8pt}  
    \renewcommand{\arraystretch}{1.2}  
    \begin{tabular}{cc}
        \toprule
        $f$ & $SNR^{min}_f$ (dB) \\
        \midrule
        7  & -7.5  \\
        8  & -10.0 \\
        9  & -12.5 \\
        10 & -15.0 \\
        11 & -17.5 \\
        12 & -20.0 \\
        \bottomrule
    \end{tabular}
    \caption{Minimum SNR for successful demodulation.} 
    \label{tab:min-snr}
\end{table}

Many algorithms designed for selecting transmission parameters rely on the \gls{snr} of packet transmissions to determine optimal settings \cite{farhad2022hadr, farhad2020enhanced, moysiadis2021extending, slabicki2018adaptive}. Given that the presented simulation model uses \gls{rss} to assess transmission success or failure, it is important for the calculated \gls{snr} to exceed the minimum threshold for successful transmissions and fall below this threshold for failed transmissions if such \gls{snr}-based algorithms are to be evaluated using the simulation framework. Consequently the \gls{snr} of a sampled transmission from ED $i$ arriving at GW $k$ is approximated as:

\begin{equation} 
\label{eq:snr}
    \hat{SNR_{ik}} = \hat{RSS_{ik}} - \frac{1}{|\mathcal{F}|} \sum_{f \in \mathcal{F}} (\eta_{f} + SNR^{min}_f)
\end{equation}

Finally, we define the \gls{ee} of ED $i$ as the number of payload bits successfully transmitted per amount of energy consumed by the ED. \gls{ee} can be calculated as follows: 

\begin{equation} \label{eq:energy-efficiency}
    EE_{i} = \frac{8 L \cdot PDR_i}{e_{p_i} \, T_i}
\end{equation}

where $L$ is the number of payload bytes, $T_i$ is the time required for transmitting a single packet, and $e_{p_i}$ is the power consumption expressed in mW using a transmission power setting~$p_i~\in~\mathcal{P}$.

\subsection{Network configuration and results}
\label{subsec:interface}
The analytical model described above forms the core of \system, enabling us to accurately model network behavior. To enhance \system's efficiency, we extend these equations to matrix operations, permitting faster computations, especially in large-scale network simulations. Based on these operations, we define the required inputs for \system, such as network configuration, and determine the simulation outputs, including key performance metrics. The simulation model, which includes $N$ end devices and $K$ gateways, operates with the following input parameters:

\begin{itemize}
    \item $X \subset \mathbb{R}^{N \times 2}$: Positions of end devices as 2-dimensional coordinates.
    \item $Y \subset \mathbb{R}^{K \times 2}$: Positions of gateways as 2-dimensional coordinates.
    \item $F \subset \mathcal{F}^N$: Assigned spreading factors for end devices.
    \item $P \subset \mathcal{P}^N$: Assigned transmission power settings for end devices.
\end{itemize}

Other network configuration settings, such as the coding rate, channel bandwidth, and packet generation rate, are considered scalar values shared across all the end devices within the simulated network. \system produces the following metrics and simulation results:

\begin{itemize}
    \item $RSS \subset \mathbb{R}^{N \times K}$: The expected \gls{rss} of transmissions from every end device at each gateway measured in dBm (ignoring signal attenuation due to flat fading).
    \item $\hat{RSS} \subset (\mathbb{R} \cup \{-\infty\})^{N \times K}$: The sampled \gls{rss} of transmissions from every end device at each gateway measured in dBm (considering signal attenuation, receiver sensitivity, and potential transmission corruption due to interference).
    \item $\hat{SNR} \subset (\mathbb{R} \cup \{-\infty\})^{N \times K}$: The sampled \gls{snr} of transmissions from every end device at each gateway measured in dB.
    \item $PDR^{GW} \subset [0,1]^{N \times K}$: The \gls{pdr} of each end device to each gateway.
    \item $PDR \subset [0,1]^N$: The overall \gls{pdr} of each end device considering all the available gateways.
    \item $EE \subset \mathbb{R}^N$: The energy efficiency of each end device measured in bits/mJ.
\end{itemize}

\subsection{ADR and backoff mechanisms} 
\system includes the ADR and backoff mechanisms proposed by Semtech \cite{semtech2016adr, semtech2023adrbackoff}, commonly used in scientific research as a comparison baseline. We have reproduced the logic of the Semtech ADR in our implementation, as detailed in Algorithm~\ref{alg:eval-adr}, and integrated the backoff variant described in Algorithm~\ref{alg:adr-backoff-implementation}. 

\subsubsection{ADR}
In the default ADR mechanism, a message buffer stores the maximum of the packet transmissions' \glspl{snr} across all gateways. Once the buffer is full, a link margin depending on the maximum \gls{snr} in the buffer and the currently assigned spreading factor is determined and a number of steps to execute are calculated (where the $\text{truncate}$ function retains the integer part of a floating point value). Depending on the number of calculated steps to execute, the spreading factor and transmission power settings are adjusted. 


\begin{algorithm}[t!]
    \caption{ADR}
    \label{alg:eval-adr}
    \begin{algorithmic}[1]
        \Require $\mathcal{F}$, $\mathcal{P}$, $\eta$
                
        \Statex\hrulefill
        
        \Procedure{init}{}
            \State initialize measurement buffer $M$ of size 20
        \EndProcedure

        \Statex\hrulefill

        \Procedure{add\_measurement}{$\hat{SNR}, f, p$}
            \State $M \xleftarrow{push} \max \hat{SNR}$
            \If{$M$ is full}
                \State $SNR_{max} \gets \max M$
                \State $SNR_{margin} \gets SNR_{max} - \eta_f - 10$
                \State $n \gets \text{truncate}(\frac{SNR_{margin}}{3})$

                \While{$n > 0 \text{ and } f > \min \mathcal{F}$}
                    \State $f \gets f-1$
                    \State $n \gets n-1$
                \EndWhile
    
                \While{$n > 0 \text{ and } p > \min \mathcal{P}$}
                    \State $p \gets p-2$
                    \State $n \gets n-1$
                \EndWhile
    
                \While{$n < 0 \text{ and } p < \max \mathcal{P}$}
                    \State $p \gets p + 2$
                    \State $n \gets n+1$
                \EndWhile
    
                \State \textbf{yield} $(f, p)$

                \State $M \xleftarrow{clear} \emptyset$
            \EndIf
        \EndProcedure

        \Statex\hrulefill

        \Procedure{reset}{}
            \State $M \xleftarrow{clear} \emptyset$
        \EndProcedure
    \end{algorithmic}
\end{algorithm}

\subsubsection{ADR backoff}
We have also implemented the ADR backoff mechanism to handle communication losses, e.g., caused by the selection of inadequate transmission parameters. Upon initialization, a counter $n_{ADR\_ACK\_CNT}$ is set to 0. This counter is increased whenever the end device attempts an uplink transmission (lines \ref{alg:line:adrbackoff-impl-reset-recv-start}-\ref{alg:line:adrbackoff-impl-reset-recv-end}). Alternatively, after $n_{ADR\_ACK\_LIMIT}$ uplink transmissions have been attempted without receiving new transmission parameters, the end device actively requests transmission parameters (lines \ref{alg:line:adrbackoff-impl-reset-req-start}-\ref{alg:line:adrbackoff-impl-reset-req-end}). If the network server receives an uplink transmission including such a request, it will reconfirm the current transmission settings with a downlink transmission (assumed to arrive at the end device), resetting the counter to 0. If transmissions still do not arrive at any gateway, after further $n_{ADR\_ACK\_DELAY}$ uplink transmissions, the transmission power will be set to the allowed maximum (lines \ref{alg:line:adrbackoff-impl-recover-start}-\ref{alg:line:adrbackoff-impl-recover-end}). Finally, after every further $n_{ADR\_ACK\_DELAY}$ transmission, the spreading factor will be increased by one step, up to the possible maximum.

\begin{algorithm}[t!]
    \caption{ADR backoff}
    \label{alg:adr-backoff-implementation}
    \begin{algorithmic}[1]
        \Require
            \Statex $\theta = (n_{ADR\_ACK\_LIMIT}, n_{ADR\_ACK\_DELAY}, \mathcal{F}, \mathcal{P})$
        \Statex\hrulefill
        \Procedure{init}{$\theta$} \Comment{Initialization} \label{alg:line:adrbackoff-impl-init-start}
            \State $n_{ACK\_CNT} \gets 0$
        \EndProcedure \label{alg:line:adrbackoff-impl-init-end}

        \Statex\hrulefill
        \Procedure{update}{} \Comment{No response received} \label{alg:line:adrbackoff-impl-update-start}
            \State $n_{ACK\_CNT} \gets n_{ACK\_CNT} + 1$
        \EndProcedure \label{alg:line:adrbackoff-impl-update-end}

        \Statex\hrulefill
        \Procedure{reset\_on\_receive}{} \Comment{Reset, upon receiving ADR} \label{alg:line:adrbackoff-impl-reset-recv-start}
            \State $n_{ACK\_CNT} \gets 0$
        \EndProcedure \label{alg:line:adrbackoff-impl-reset-recv-end}

        \Statex\hrulefill
        \Procedure{reset\_on\_request}{} \Comment{Reset, with requesting ADR} \label{alg:line:adrbackoff-impl-reset-req-start}
            \If{$n_{ACK\_CNT} > n_{ADR\_ACK\_LIMIT}$}
                \State $n_{ACK\_CNT} \gets 0$
            \EndIf
        \EndProcedure \label{alg:line:adrbackoff-impl-reset-req-end}

        \Statex\hrulefill
        \Procedure{recover}{$f_{current},p_{current}$} \Comment{Adjust transmission parameters} \label{alg:line:adrbackoff-impl-recover-start}
            \State $f_{new} \gets f_{current}$
            \State $p_{new} \gets p_{current}$
            \If{$n_{ADR\_ACK\_CNT} = n_{ADR\_ACK\_LIMIT} + n_{ADR\_ACK\_DELAY}$}
                \State $p_{new} \gets \text{max}(\mathcal{P})$
            \ElsIf{$n_{ADR\_ACK\_CNT} > n_{ADR\_ACK\_LIMIT} + n_{ADR\_ACK\_DELAY}$}
                \If{$n_{ADR\_ACK\_CNT} - n_{ADR\_ACK\_LIMIT} \equiv 0\ (\text{mod }n_{ADR\_ACK\_DELAY})$}
                    \State $f_{new} \gets \text{min}(f_{current}+1, \text{max}(\mathcal{F}))$
                \EndIf
            \EndIf
            \State \textbf{return} $f_{new}, p_{new}$
             \label{alg:line:adrbackoff-impl-recover-end}
        \EndProcedure 
    \end{algorithmic}
\end{algorithm}

\subsection{Implementation}
\system has been implemented leveraging matrix operations using the well-known NumPy\footnote{\url{https://numpy.org/}} library, efficiently applying caching techniques to avoid redundant calculations. The source code of \system and the default ADR and backoff mechanisms have been made publicly available \cite{repository_fast-lora} 
as a fully open-source Python package. This not only ensures transparency and reproducibility but also permits seamless integration with common libraries that can be used to optimize LoRa networks \cite{GymDocum55, RLlibInd4, PettingZ20, StableBa63}.

\section{Evaluation setup}
\label{sec:setup}
This section describes the configuration setup to be evaluated, detailing the baselines used, network configuration, performance metrics, and evaluation scenarios. During this evaluation, we focus on congested multi-gateway networks to cover all the aspects of the analytical model behind \system (e.g., interference among EDs). We simulate different distances between EDs and GWs to verify the correct calculation of signal strength combined with receiver sensitivity. In this evaluation, we assume homogeneous receiver sensitivities across all the devices in the network, meaning all EDs are modeled with identical sensitivity, power consumption, and transmission settings, as is common in other simulators \cite{bor2016lora, slabicki2022flora, magrin2016network}. However, some real-world deployments may exhibit heterogeneity in these aspects, which could impact simulation accuracy in scenarios where such variability significantly influences communication reliability. Furthermore, by simulating a large number of EDs, we ensure correct interference calculations. Including multiple active GWs within a network allows us to verify the accuracy of reception calculations at the network server level. The network parameters selected for the validation process are listed in Table~\ref{tab:validation_parameters}. All simulations have been conducted on an Intel Core i5-8265U processor with 8~GB of RAM.

\subsection{Metrics}
To evaluate \system, we employ key performance metrics, such as the Packet Delivery Ratio (PDR) and Energy Efficiency (EE). The PDR is defined as the ratio of successfully received packets to the total number of packets transmitted by ED $i$:

\begin{equation} 
\label{eq:pdr-flora}
    PDR_i = \frac{n_i^{rx}}{n_i^{tx}}
\end{equation}

where $n_i^{rx}$ is the number of packets received by any gateway in the network, and $n_i^{tx}$ is the total number of packets transmitted by ED $i$. Similarly, EE is defined as the number of payload bits successfully transmitted per unit of energy consumed by the ED, as shown in Eq.~\eqref{eq:energy-efficiency}.

\subsection{Baselines} 
\system correctness and accuracy are validated using FLoRa, a well-known discrete event-based simulation framework for LoRa networks \cite{slabicki2022flora}. While \system computes these metrics directly based on statistical models, FLoRa simulates individual packet transmissions throughout different layers of the communication protocol and network components. With this comparison, we ensure that \system provides accurate network performance estimations and results with significantly reduced computational overhead. Unlike FLoRa and other simulators \cite{magrin2016network, slabicki2022flora}, \system focuses on efficiently modeling key performance metrics, making it particularly suitable for large-scale networks and dynamic environments. 

To achieve a fair comparison, we carefully reviewed the implementation of FloRa and appropriately configured the random packet generation and energy consumption calculations. FLoRa models the time between ED packet generation using an exponential distribution. To maintain compatibility with the underlying simulation framework, constraints were introduced for different spreading factors to prevent potential runtime errors \cite{slabicki2022flora}. As a result, these constraints cause the time between packet generations to deviate from a strictly exponential distribution. These constraints are only required in edge cases, and the simulation framework generally allows shorter delays without issues. Since the assumption of an exponential inter-arrival time is essential to accurately model interference probability, we opted to disable the original constraints while implementing an error-handling mechanism to manage the exceptional cases where constraints were required. 

In this work, \system focuses on energy consumption analysis and calculation of energy consumption in uplink packets. FLoRa estimates energy consumption using predefined table values without a specified data source. Upon review, we observed that some transmission power levels were assigned identical consumption values (e.g., 24 mA for 2–4 dBm, 25 mA for 5–8 dBm), which deviates from experimentally measured values \cite{liando2019known}. To improve the accuracy of the validation process, we set the power consumption values to better reflect the measured data. We also excluded the energy consumption of downlink communications. It is important to note that the validation focuses on scenarios that match the assumptions made by \system, such as homogeneous device behavior, uplink-only communication, and static traffic patterns. As such, the accuracy results reported here are specific to this class of scenarios, for which \system is designed to serve as a reliable and efficient approximation.

\begin{table}[ht!]
    \centering
    \setlength{\tabcolsep}{8pt}  
    \renewcommand{\arraystretch}{1.2}  
    
    \begin{tabular}{lc}
        \toprule
        Parameter & Value \\
        \midrule
         $\mathcal{P}$ & $\{2, 4, 6, 8, 10, 12, 14, 16\}~\text{dBm}$ \\
         $\mathcal{F}$ & $\{7, 8, 9, 10, 11, 12\}$ \\
         $r$ & $1$ \\
         $B$ & 125 kHz \\
         $n_{pr}$ & 8 \\
         $CRC$ & 1 \\
         $H$ & 1 \\
         $DE$ & 0 ($f$=7-10), 1 ($f$=11-12) \\
         $\overline{PL}(d_{0})$ & 127.41 dBm \\
         $d_0$ & 40 m \\
         $\gamma$ & 2.08 \\
         $\lambda$ & $0.001~s^{-1}$ \\
         $\sigma$ & 3.57 \\
         $\omega_{f_{i}f_{j}}$ & See Table \ref{tab:sir-thresholds} \\
         $\eta_{f_{i}}$ & See references \cite{semtech2017sxdatasheet, slabicki2022flora} \\
         $e_{p_i}$ & See reference \cite{liando2019known} \\
        \bottomrule
    \end{tabular}
    \caption{Network parameters employed during the evaluation \cite{bor2016lora, slabicki2022flora, semtech2017sxdatasheet}.}  
    \label{tab:validation_parameters}
\end{table}

\begin{table}[ht!]
    \centering
    \setlength{\tabcolsep}{8pt}  
    \renewcommand{\arraystretch}{1.2}  

    \begin{tabular}{ccccccc}
        \toprule
        \diagbox{$f_{i}$}{$f_{j}$} & 7 & 8 & 9 & 10 & 11 & 12 \\
        \midrule
        7  & 6  & -8  & -9  & -9  & -9  & -9  \\
        8  & -11  & 6  & -11  & -12  & -13  & -13  \\
        9  & -15  & -13  & 6  & -13  & -14  & -15  \\
        10 & -19  & -18  & -17  & 6  & -17  & -18  \\
        11 & -22  & -22  & -21  & -20  & 6  & -20  \\
        12 & -25  & -25  & -25  & -24  & -23  & 6  \\
        \bottomrule
    \end{tabular}
    \caption{Signal-to-interference ratio thresholds $\omega_{f_{i}f_{j}}$ in dB for successful transmission decoding \cite{amichi2020joint, slabicki2022flora, toro2021modeling, xu2022x} }
    \label{tab:sir-thresholds}
\end{table}


\section{Performance evaluation results}
\label{sec:results}
This section describes the experimental scenarios used for validating \system and the methodology applied to carry out the experiments. We evaluate using three different experiments. The first two analyze performance metrics accuracy in simple and dense networks with a large number of devices, where factors such as interference may affect transmission reliability. In the third experiment, we assess the computation efficiency and simulation time of a wide range of network sizes and different configurations.

\subsection{End device to gateway communication}
In the first experiment, we validated the basic functionality of \system by comparing the PDR and EE calculations for a network with only one end device and one gateway. This extremely simplified setup allows for a direct comparison of \system and FLoRa performance in a controlled environment. The experiment was conducted by varying the ED-GW distance, considering values of 20, 50, and 100 m. The spreading factors and transmission power values were configured according to the values specified in Table \ref{tab:validation_parameters}. This setup allowed us to test the impact of these parameters on the \gls{pdr} and \gls{ee} performance metrics. The simulations were run for 120 hours using FLoRa, while \system computed the metrics without the need to simulate individual packets due to the statistical nature of its underlying mathematical model. In FloRa, ADR mechanisms were disabled to maintain the initially configured spreading factor and transmission power settings throughout the simulation. The results were then compared to assess the accuracy and efficiency of \system, directly comparing it with FLoRa.


\begin{figure*}[ht!]
    \centering    
    \subfloat[]{%
        \includegraphics[width=0.24\textwidth]{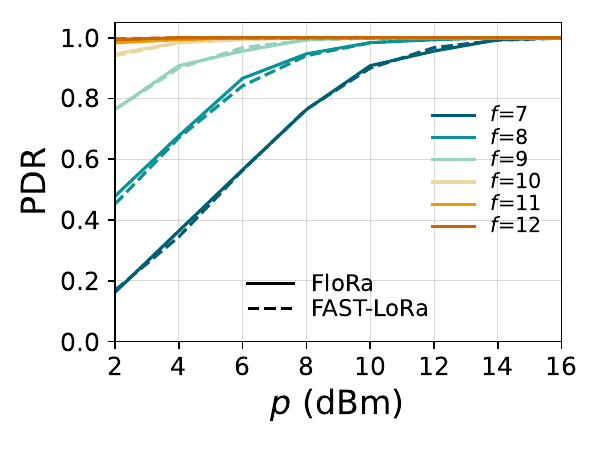}%
        \label{fig:model-verification-single-pdr-tp}%
    }
    \hfill
    \subfloat[]{%
        \includegraphics[width=0.24\textwidth]{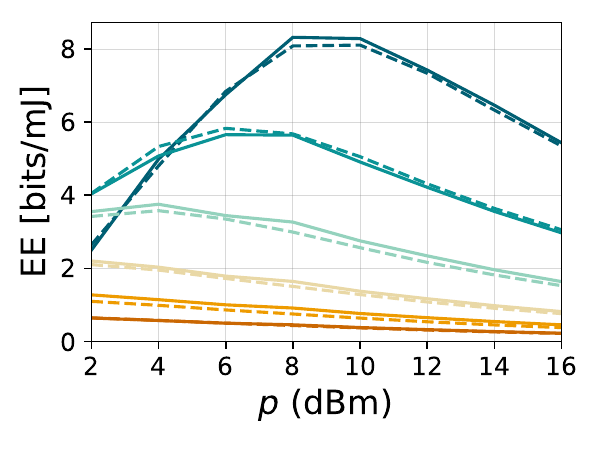}%
        \label{fig:model-verification-single-ee-tp}%
    }
    \hfill
    \subfloat[]{%
        \includegraphics[width=0.24\textwidth]{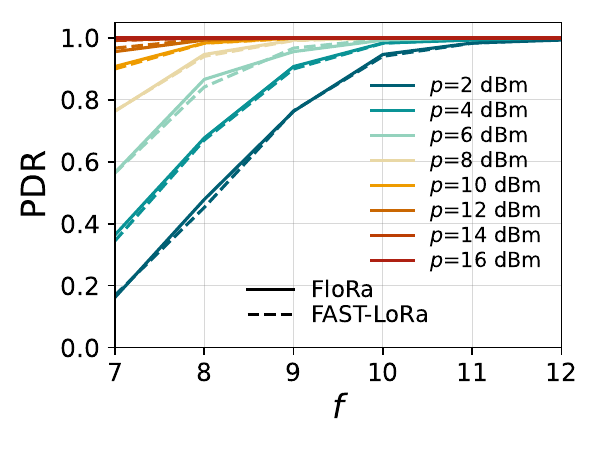}%
        \label{fig:model-verification-single-pdr-sf}%
    }
    \hfill
    \subfloat[]{%
        \includegraphics[width=0.24\textwidth]{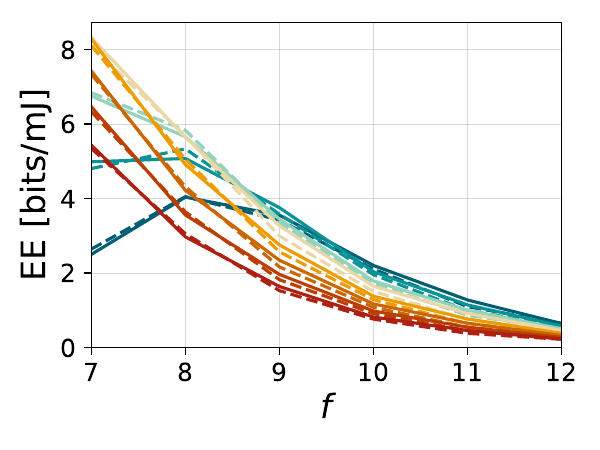}%
        \label{fig:model-verification-single-ee-sf}%
    }
    \caption{Comparison of the PDR and EE obtained with FloRa and \system{} considering different values of \(f\) (a, b) and different values of \(p\) (c, d) for an ED-GW distance of 50~m.}
    \label{fig:model-verification-single}
\end{figure*}

\begin{figure*}[t!]
    \centering    
    \subfloat[]{%
        \includegraphics[width=0.48\textwidth]{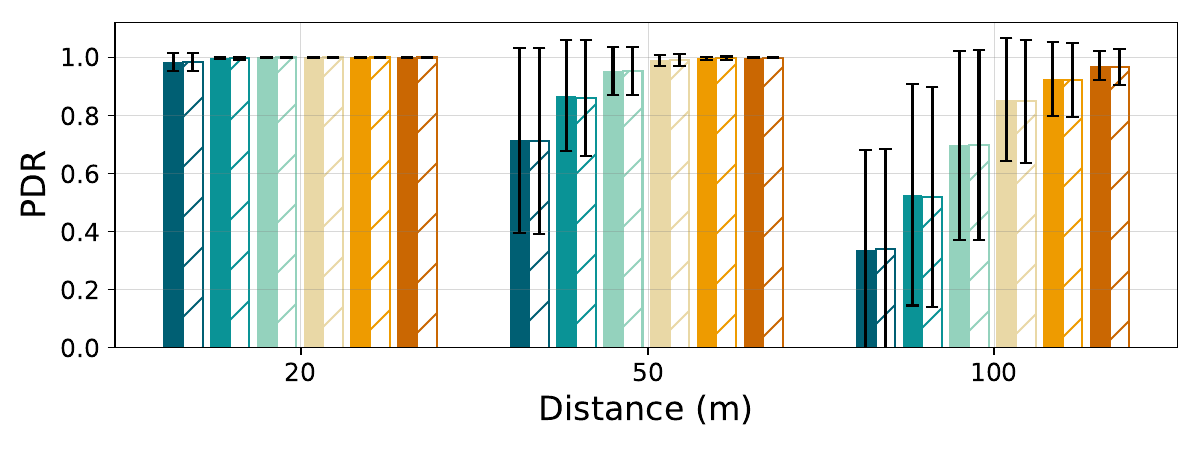}%
        \label{fig:model-verification-single-pdr-dist}%
    }
    \hfill
    \subfloat[]{%
        \includegraphics[width=0.48\textwidth]{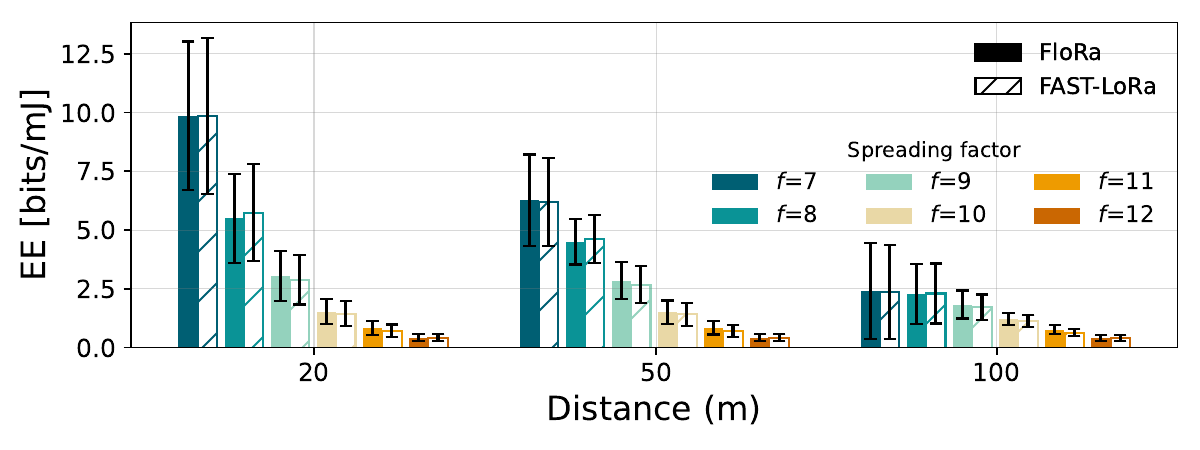}%
        \label{fig:model-verification-single-ee-dist}%
    }
    \caption{Comparison of the PDR (a) and EE (b) of FloRa and \system{} as a function of ED-GW distance for different values of \(f\).}
    \label{fig:model-verification-single-bars}
\end{figure*}

The results of this comparison are illustrated in Fig.~\ref{fig:model-verification-single} and Fig.~\ref{fig:model-verification-single-bars}. The behavior of the PDR and EE for different values of spreading factor ($f$) and transmission power ($p$) are shown in Fig.~\ref{fig:model-verification-single}. In the first plot, where the PDR is shown as a function of $f$, both \system and FLoRa exhibit similar performance, and the PDR increases with higher spreading factors. This behavior is expected, as a higher spreading factor leads to better reception, albeit at the cost of longer transmission times. The accuracy of \system's PDR calculations is confirmed, with results closely matching those of FLoRa. In the second plot, where the PDR is plotted as a function of $p$, both models show that increasing transmission power improves the PDR. However, \system computes these values significantly faster, providing near-instantaneous results compared to FLoRa's more computationally expensive approach, which requires simulating each packet. This will be further and quantitatively analyzed in Section~\ref{subsec:computation-speed-evaluation}. The third plot, which depicts EE as a function of $f$, shows that EE decreases with higher spreading factors, as expected due to longer transmission times. \system produces results comparable to FLoRa but much more efficiently, confirming the model's ability to calculate energy efficiency quickly and accurately. Finally, in the fourth plot, where EE is shown as a function of $p$, we observe that EE decreases as transmission power increases, which aligns with the expected behavior of a real-world system. \system handles this calculation in a fraction of the time that FLoRa requires, further highlighting its suitability for rapid simulations and network evaluations. The results shown in Fig.~\ref{fig:model-verification-single-bars} provide the same analysis as Fig.~\ref{fig:model-verification-single} but consider the average values across different transmission powers as a function of the distance. The first plot illustrates how the PDR decreases with increasing distance, with \system and FLoRa following similar trends. The second plot presents EE behavior across different distances, confirming the expected decline as distance increases. As observed previously, \system provides results comparable to FLoRa but with significantly reduced computation time, making it highly suitable for large-scale simulations.

To assess the simulation and accuracy quality of \system, two key error metrics were employed: the Mean Absolute Error (MAE) and the Standard Deviation of Error (SDE). These metrics help quantify how well \system approximates the performance metrics obtained from FLoRa simulations. First, we present the results from Table~\ref{tab:model-verification-single-sf}, showing the MAE and SDE values for the PDR and EE under different spreading factor ($f$) settings. This table calculates the MAE and SDE by averaging the results over the different transmission power values considered in this work. In this table, we observe how the performance of \system compares to FLoRa as the spreading factor is varied for different ED-GW distances. For example, at a distance of 20 meters, the MAE for the PDR significantly decreases as the spreading factor increases, reaching zero values at $f$=12. This suggests that the PDR of \system becomes more accurate as the spreading factor increases. Similarly, the MAE for EE follows the same trend, with fewer errors at higher spreading factor values. The SDE values show similar behavior, with low values indicating that the error is consistent and stable across different spreading factor settings. However, when the ED-GW distance increases, we notice a corresponding increase in the MAE values for the PDR and EE. This is expected because as the distance increases, the signal strength diminishes, leading to greater variability in the performance metrics. Despite this, the MAE and SDE still show a clear trend where higher spreading factor values continue to improve the accuracy of the calculations, albeit with slightly larger errors at longer distances. In summary, Table~\ref{tab:model-verification-single-sf} demonstrates that as the ED-GW distance increases, \system becomes slightly less accurate, but the general trend remains that higher spreading factor values improve the MAE and SDE, indicating that \system is still able to provide consistent and reliable results under varying conditions.

Table~\ref{tab:model-verification-single-tp} presents the MAE and SDE for the PDR and EE with varying transmission power ($p$) settings. Like the previous table, these values are averaged over different spreading factor settings. This table provides insights into how transmission power affects \system accuracy and consistency of \system performance. For the PDR metric, the MAE decreases as $p$ increases, which is expected, as higher transmission power typically leads to better signal reception and, consequently, more accurate predictions. At a distance of 20 meters, for instance, the MAE for the PDR approaches zero as the $p$ increases, showing that the model is increasingly accurate with higher transmission power settings. The SDE values for the PDR also decrease, further indicating that \system provides stable and consistent results across different transmission power levels. However, as the ED-GW distance increases (e.g., at 50 or 100 meters), we notice that the MAE for the PDR begins to rise. This is expected because, as distance increases, signal strength diminishes, leading to varying network performance and less accurate values. Even so, the MAE for the PDR still decreases with higher transmission power values, indicating that increasing transmission power helps mitigate the effects of increased distance and signal degradation. Similarly, the SDE values increase slightly with distance, reflecting a greater variability in the computed PDR values as distance grows. However, they remain relatively low for higher transmission power values. For EE, a similar behavior is observed. The MAE decreases as the transmission power increases, and the SDE values remain relatively low, reflecting the model’s reliability in predicting energy efficiency across different transmission power settings. As with the PDR, at greater distances (e.g., at 100 meters), the MAE for EE is slightly higher, but the increase is not as pronounced as for the PDR. At higher transmission power values (e.g., $p$=16), both the PDR and EE values from \system are very close to the FLoRa results, confirming the robustness of \system in accurately estimating network performance, even as both transmission power and distance vary.

\begin{table}[t!]
    \centering
    \small
    \setlength{\tabcolsep}{6pt}  
    \renewcommand{\arraystretch}{1.2}  
    
    \begin{tabular}{llcccc}
        \toprule
        \multirow{2}{*}{Distance (m)} & \multirow{2}{*}{$f$} & \multicolumn{2}{c}{PDR [$\boldsymbol{\times 10^{-2}}$]} & \multicolumn{2}{c}{EE (bits/mJ)} \\
        \cmidrule(lr){3-4} \cmidrule(lr){5-6}
        & & MAE & SDE & MAE & SDE \\
        \midrule
        \multirow{6}{*}{20} & 7  & 0.122 & 0.214 & 0.201 & 0.219 \\
                            & 8  & 0.089 & 0.168 & 0.230 & 0.177 \\
                            & 9  & 0.017 & 0.037 & 0.172 & 0.045 \\
                            & 10 & 0.001 & 0.002 & 0.085 & 0.022 \\
                            & 11 & 0.000 & 0.000 & 0.131 & 0.034 \\
                            & 12 & 0.000 & 0.000 & 0.008 & 0.007 \\
        \midrule
        \multirow{6}{*}{50} & 7  & 0.630 & 0.855 & 0.143 & 0.126 \\
                            & 8  & 0.846 & 1.059 & 0.111 & 0.074 \\
                            & 9  & 0.295 & 0.493 & 0.162 & 0.052 \\
                            & 10 & 0.132 & 0.236 & 0.086 & 0.024 \\
                            & 11 & 0.057 & 0.114 & 0.130 & 0.033 \\
                            & 12 & 0.048 & 0.109 & 0.008 & 0.008 \\
        \midrule
        \multirow{6}{*}{100} & 7  & 0.574 & 0.740 & 0.064 & 0.087 \\
                             & 8  & 1.347 & 1.631 & 0.079 & 0.087 \\
                             & 9  & 0.595 & 0.735 & 0.117 & 0.049 \\
                             & 10 & 0.820 & 1.226 & 0.081 & 0.034 \\
                             & 11 & 0.751 & 1.237 & 0.122 & 0.031 \\
                             & 12 & 0.614 & 1.119 & 0.010 & 0.007 \\
        \midrule
        All & All & 0.385 & 0.811 & 0.108 & 0.129 \\
        \bottomrule
    \end{tabular}
    \caption{MAE and SDE values for PDR and EE across different spreading factor ($f$) settings. The results are averaged over different transmission power ($p$) configurations and ED-GW distances between the ED and GW.} 
    \label{tab:model-verification-single-sf}
\end{table}

\begin{table}[t!]
    \centering
    \small
    \setlength{\tabcolsep}{6pt}  
    \renewcommand{\arraystretch}{1.2}  
    
    \begin{tabular}{llcccc}
        \toprule
        \multirow{2}{*}{Distance (m)} & \multirow{2}{*}{$p$} & \multicolumn{2}{c}{PDR [$\boldsymbol{\times 10^{-2}}$]} & \multicolumn{2}{c}{EE (bits/mJ)} \\
        \cmidrule(lr){3-4} \cmidrule(lr){5-6}
        & & MAE & SDE & MAE & SDE \\
        \midrule
        \multirow{8}{*}{20} & 2  & 0.101 & 0.145 & 0.201 & 0.246 \\
                            & 4  & 0.169 & 0.245 & 0.199 & 0.245 \\
                            & 6  & 0.011 & 0.019 & 0.156 & 0.194 \\
                            & 8  & 0.021 & 0.044 & 0.167 & 0.140 \\
                            & 10 & 0.003 & 0.006 & 0.113 & 0.109 \\
                            & 12 & 0.000 & 0.001 & 0.109 & 0.093 \\
                            & 14 & 0.000 & 0.000 & 0.087 & 0.077 \\
                            & 16 & 0.000 & 0.000 & 0.069 & 0.064 \\
        \midrule
        \multirow{8}{*}{50} & 2  & 0.728 & 1.069 & 0.095 & 0.108 \\
                            & 4  & 0.634 & 0.721 & 0.142 & 0.154 \\
                            & 6  & 0.685 & 1.095 & 0.094 & 0.109 \\
                            & 8  & 0.156 & 0.232 & 0.144 & 0.110 \\
                            & 10 & 0.170 & 0.280 & 0.123 & 0.115 \\
                            & 12 & 0.244 & 0.415 & 0.096 & 0.087 \\
                            & 14 & 0.033 & 0.052 & 0.088 & 0.078 \\
                            & 16 & 0.027 & 0.056 & 0.070 & 0.065 \\
        \midrule
        \multirow{8}{*}{100} & 2  & 0.997 & 1.466 & 0.044 & 0.051 \\
                             & 4  & 1.034 & 1.433 & 0.059 & 0.076 \\
                             & 6  & 1.741 & 1.596 & 0.107 & 0.097 \\
                             & 8  & 0.569 & 0.648 & 0.084 & 0.073 \\
                             & 10 & 0.402 & 0.496 & 0.100 & 0.101 \\
                             & 12 & 0.988 & 1.402 & 0.108 & 0.062 \\
                             & 14 & 0.386 & 0.579 & 0.065 & 0.064 \\
                             & 16 & 0.150 & 0.212 & 0.063 & 0.059 \\
        \midrule
        All & All & 0.385 & 0.811 & 0.108 & 0.129 \\
        \bottomrule
    \end{tabular}
    \caption{MAE and SDE values for the PDR and EE with varying transmission power ($p$) settings. The results are averaged over different spreading factor ($f$) configurations and ED-GW distances.}
    \label{tab:model-verification-single-tp}
\end{table}

\subsection{Realistic dense networks}
\begin{figure}[ht!]
    \centering
    \includegraphics[width=0.75\linewidth]{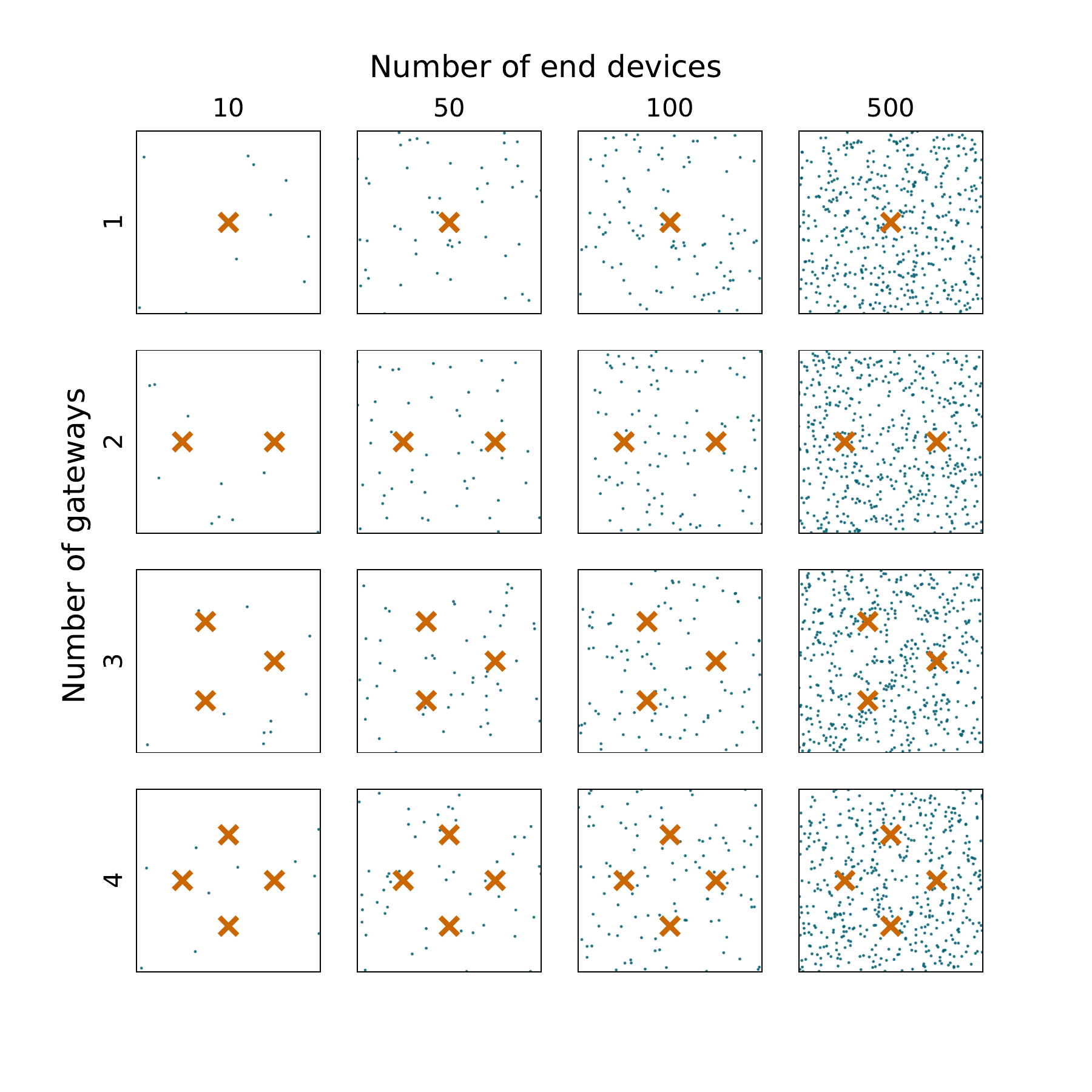}
    \caption{Different network configurations including end devices (blue points) and gateways (orange crosses) within the 1x1 km area used for evaluation.}
    \label{fig:model-verification-multi-placement}
\end{figure}

In the second validation step, we extended the evaluation to a more complex network scenario involving multiple end devices (EDs) and gateways (GWs) to assess \system's performance in a real-world-like network with interference and reception calculations. In this setup, different network configurations were simulated, considering 10, 50, 100, and 500 EDs, along with 1 to 4 GWs, to evaluate how well \system handles interference and reception when combining data from multiple gateways. For each combination of end devices and gateways, a total of 5 network configurations were generated to ensure randomness without introducing unintentional bias. The spreading factor and transmission power settings were randomly assigned to the end devices, sampled from all possible configurations (as shown in Table~\ref{tab:validation_parameters}), ensuring broad coverage of different network configurations. Additionally, the positions of the end devices were randomly assigned by sampling x- and y-coordinates from a uniform distribution, limiting both coordinates between 0~m and 1000~m. This resulted in diverse network topologies, as seen in Fig.~\ref{fig:model-verification-multi-placement}, where the gateways are represented in orange and the end devices in blue. 

In this scenario, both \system and FLoRa were used to calculate the PDR and EE for each end device. The FLoRa simulations ran for 5 days, simulating approximately 432 packet transmissions per ED using the parameters specified in Table~\ref{tab:validation_parameters}, resulting in minor statistical fluctuations that were considered acceptable for validation purposes. ADR mechanisms were disabled to ensure that the initially assigned spreading factor and transmission power settings were maintained throughout the simulations. Due to the statistical nature of \system, it calculated the PDR and EE through faster simulations, avoiding the need for packet-based simulations. The MAE and SDE of the PDR and EE were calculated across all EDs and simulations for each network configuration, and the results are presented in Table~\ref{tab:model-verification-multi-pdr-mae} and Table~\ref{tab:model-verification-multi-ee-mae}, respectively. From Table~\ref{tab:model-verification-multi-pdr-mae}, we can observe that as the number of gateways and end devices increases, the MAE and SDE tend to rise slightly. As we increase the number of end devices and gateways, the MAE for the PDR shows slight increases, reaching a value of 1.044$\times10^{-2}$ for the MAE and 1.579$\times10^{-2}$ for the SDE in networks with 500 EDs and 4 GWs. Despite this increase, both error metrics remain within an acceptable range, which reflects that \system's PDR values are accurate in complex network configurations. Overall, \system's PDR results have an average MAE of 0.940$\times10^{-2}$ and an average SDE of 1.498$\times10^{-2}$ across all network configurations. Similarly, Table~\ref{tab:model-verification-multi-ee-mae} shows the MAE and SDE values for EE, with the trend being quite similar to that observed for the PDR. Both the MAE and SDE for EE are small across all network configurations, with an average MAE of 0.038 bits/mJ and an average SDE of 0.068 bits/mJ, demonstrating the consistency of \system’s EE calculations. Even with higher network sizes, the error remains low, with slight increases in the MAE and SDE as network complexity grows. For instance, in networks with 500 EDs and 4 GWs, the MAE increases slightly to 0.055 bits/mJ and the SDE to 0.078  bits/mJ. In conclusion, Table~\ref{tab:model-verification-multi-pdr-mae} and Table~\ref{tab:model-verification-multi-ee-mae} show the accuracy of \system and the consistency of the results across different network configurations, with MAE and SDE values remaining low enough to be considered reliable for practical use in network optimization. These results confirm that \system is capable of providing efficient and precise PDR and EE values, even in complex and large-scale networks.

\begin{table}[ht!]
    \centering
    \small
    \begin{tabular}{ccccc}
        \toprule
        \diagbox{EDs}{GWs} & 1 & 2 & 3 & 4 \\
        \midrule
        10  & 0.670 (1.182) & 0.960 (1.509) & 1.067 (1.730) & 0.765 (1.390) \\
        50  & 0.718 (1.207) & 0.671 (1.109) & 0.852 (1.426) & 0.857 (1.404) \\
        100 & 0.645 (1.174) & 0.759 (1.237) & 0.820 (1.367) & 0.935 (1.451) \\
        500 & 0.868 (1.464) & 1.040 (1.590) & 1.002 (1.526) & 1.044 (1.579) \\
        \bottomrule
    \end{tabular}
    \caption{MAE ($\times 10^{-2}$) and SDE (in brackets) values for the PDR calculated across all EDs and network configurations, with varying numbers of EDs and GWs.}
    \label{tab:model-verification-multi-pdr-mae}
\end{table}

\begin{table}[ht!]
    \centering
    \small
    \begin{tabular}{ccccc}
        \toprule
        \diagbox{EDs}{GWs} & 1 & 2 & 3 & 4 \\
        \midrule
        10  & 0.022 (0.035) & 0.034 (0.051) & 0.040 (0.065) & 0.042 (0.072) \\
        50  & 0.021 (0.036) & 0.039 (0.070) & 0.050 (0.071) & 0.049 (0.073) \\
        100 & 0.018 (0.037) & 0.030 (0.050) & 0.042 (0.064) & 0.055 (0.074) \\
        500 & 0.022 (0.042) & 0.038 (0.061) & 0.048 (0.071) & 0.055 (0.078) \\
        \bottomrule
    \end{tabular}
    \caption{MAE and SDE (in brackets) values for EE [bits/mJ] calculated across all EDs and network configurations, with varying numbers of EDs and GWs.}
    \label{tab:model-verification-multi-ee-mae}
\end{table}

\subsection{Computational performance}
\label{subsec:computation-speed-evaluation}
\begin{figure}[ht!]
    \centering
    \includegraphics[width=0.75\linewidth]{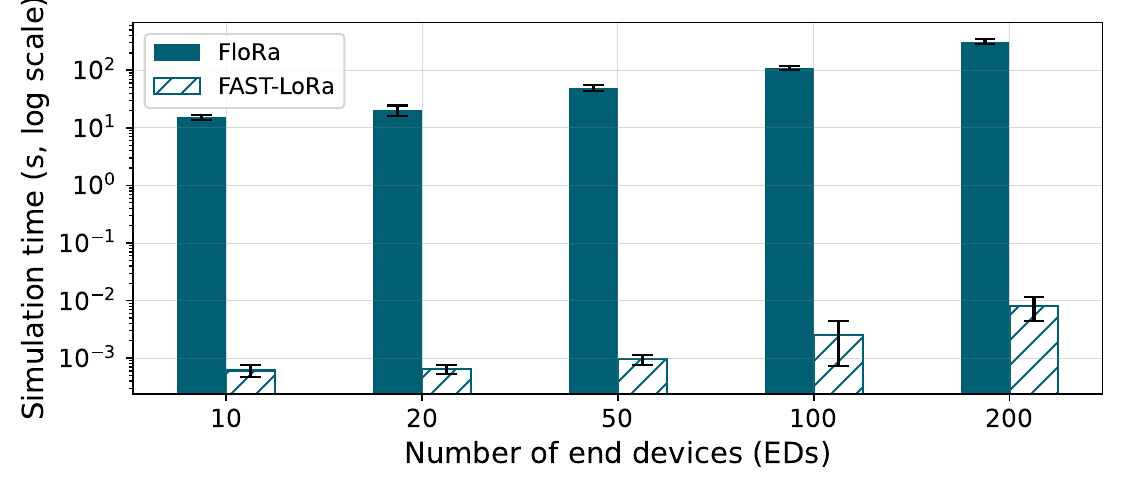}
    \caption{Comparison of the average execution time of FloRa and FAST-LoRa as the number of end devices increases. Results are averaged over simulations with 1 to 5 gateways and shown on a logarithmic scale.}
    \label{fig:average_execution_time}
\end{figure}

After evaluating the performance of both approaches in terms of network-related metrics, this section focuses on the computational performance of FloRa and \system. Specifically, we compare their times when simulating networks with increasing numbers of end devices and gateways, to assess the computational cost and scalability of each framework. As Figure \ref{fig:average_execution_time} shows, \system consistently achieves significantly lower simulation times than FloRa across all network sizes. For instance, with only 10 EDs, \system completes a simulation in approximately 0.6 milliseconds, while FloRa takes over 15 seconds. This gap becomes even more pronounced as the network grows: with 200 EDs, the scenario with the highest end device density evaluated, FloRa exceeds 5 minutes, whereas \system remains under 8 milliseconds. This results in an acceleration of up to three orders of magnitude in favor of \system. This significant improvement is mainly due to \system’s core design, which relies on an analytical model that obtains network performance metrics without the computational burden associated with simulating complex network behaviors that are not directly relevant to transmission parameter selection.

\section{Conclusions}
\label{sect:conc_fw}
In this work, we have introduced \system, an efficient simulation framework designed to enable fast and efficient optimizations for LoRaWAN networks. This is achieved through a statistical approach that focuses on computing core performance metrics rather than including multiple layers of communication protocols. To enable this, \system applies a set of modeling simplifications, such as assuming a single-channel setup, replacing packet-level interactions with analytical models, and computing gateway receptions through matrix operations. By doing so, \system reduces computational complexity, making it particularly useful for online optimization techniques or machine learning methods requiring many iterations. A key aspect of this work is its reproducibility. The \system framework as well as the experiments and results provided in this paper are publicly available and, therefore, fully reproducible. Our evaluation demonstrates that \system achieves comparable accuracy to the well-established FLoRa simulator, with a Mean Absolute Error (MAE) of 0.940 $\times 10^{-2}$ for the Packet Delivery Ratio (PDR) and 0.040 bits/mJ for Energy Efficiency (EE) in complex scenarios with interference and multi-gateway reception. Moreover, in terms of computational performance, \system completes simulations in a fraction of the time required by FloRa. In the network with the highest density, with 200 end devices and 5 gateways, \system executes in under 8 milliseconds, compared to over 5 minutes for FloRa. While this approach significantly reduces computational overhead, it also abstracts certain LoRaWAN-specific mechanisms, such as downlink communication and detailed medium access interactions. However, simulating complex network behaviors that are not directly relevant in scenarios focused on transmission parameter optimization can introduce computational overhead. \system enables the simulation of large-scale LoRaWAN networks with numerous devices and gateways, making it particularly suitable for fast optimizations of transmission parameters. However, while \system provides an efficient solution for many scenarios, some limitations exist due to its simplifications. For instance, the model assumes a homogeneous radio propagation model and focuses only on uplink communications, which may not fully capture the behavior of real-world, heterogeneous networks or downlink communication aspects. Therefore, rather than aiming for a replacement for detailed packet-level simulators, \system is intended as a lightweight and practical alternative for scenarios where its modeling assumptions hold, such as static uplink traffic and homogeneous device behavior. Future work will focus on addressing these limitations and extending the framework to support a more comprehensive analysis of LoRaWAN networks.

\section*{Acknowledgments}
\begin{minipage}{\linewidth}
This work was supported by the grant TED2021-129336B-I00, funded by MCIN/AEI/10.13039/501100011033 and by the European Union NextGenerationEU/PRTR; by PID2023-148214OB-C21, funded by MICIU/AEI/10.13039/501100011033 and by FEDER, EU; and by the grant PCI2024-153485, funded by MICIU/AEI/10.13039/501100011033 and the European Union. Minor edits to this manuscript for clarity and correctness were supported using OpenAI's ChatGPT.
\end{minipage}

\bibliographystyle{unsrt}  
\bibliography{references}

\begin{thebibliography}{10}

\bibitem{gkotsiopoulos2021performance}
Panagiotis Gkotsiopoulos, Dimitrios Zorbas, and Christos Douligeris.
\newblock Performance determinants in lora networks: A literature review.
\newblock {\em IEEE Communications Surveys \& Tutorials}, 23(3):1721--1758, 2021.

\bibitem{sundaram2019survey}
Jothi Prasanna~Shanmuga Sundaram, Wan Du, and Zhiwei Zhao.
\newblock A survey on lora networking: Research problems, current solutions, and open issues.
\newblock {\em IEEE Communications Surveys \& Tutorials}, 22(1):371--388, 2019.

\bibitem{benkahla2019enhanced}
Norhane Benkahla, Hajer Tounsi, SONG Ye-Qiong, and Mounir Frikha.
\newblock Enhanced adr for lorawan networks with mobility.
\newblock In {\em 2019 15th International Wireless Communications \& Mobile Computing Conference (IWCMC)}, pages 1--6. IEEE, 2019.

\bibitem{farhad2020enhanced}
Arshad Farhad, Dae-Ho Kim, Santosh Subedi, and Jae-Young Pyun.
\newblock Enhanced lorawan adaptive data rate for mobile internet of things devices.
\newblock {\em Sensors}, 20(22):6466, 2020.

\bibitem{moysiadis2021extending}
Vasileios Moysiadis, Thomas Lagkas, Vasileios Argyriou, Antonios Sarigiannidis, Ioannis~D Moscholios, and Panagiotis Sarigiannidis.
\newblock Extending adr mechanism for lora enabled mobile end-devices.
\newblock {\em Simulation Modelling Practice and Theory}, 113:102388, 2021.

\bibitem{anwar2021rm}
Khola Anwar, Taj Rahman, Asim Zeb, Inayat Khan, Mahdi Zareei, and Cesar Vargas-Rosales.
\newblock Rm-adr: Resource management adaptive data rate for mobile application in lorawan.
\newblock {\em Sensors}, 21(23):7980, 2021.

\bibitem{jouhari2023deep}
Mohammed Jouhari, Khalil Ibrahimi, Jalel~Ben Othman, and El~Mehdi Amhoud.
\newblock Deep reinforcement learning-based energy efficiency optimization for flying lora gateways.
\newblock In {\em ICC 2023-IEEE International Conference on Communications}, pages 6157--6162. IEEE, 2023.

\bibitem{muthanna2022deep}
Mohammed Saleh~Ali Muthanna, Ammar Muthanna, Ahsan Rafiq, Mohammad Hammoudeh, Reem Alkanhel, Stephen Lynch, and Ahmed~A Abd El-Latif.
\newblock Deep reinforcement learning based transmission policy enforcement and multi-hop routing in qos aware lora iot networks.
\newblock {\em Computer Communications}, 183:33--50, 2022.

\bibitem{mhatre2022dynamic}
Jui Mhatre and Ahyoung Lee.
\newblock Dynamic reinforcement learning based scheduling for energy-efficient edge-enabled lorawan.
\newblock In {\em 2022 IEEE International Performance, Computing, and Communications Conference (IPCCC)}, pages 412--413. IEEE, 2022.

\bibitem{yazid2022reinforcement}
Yassine Yazid, Antonio Guerrero-Gonz{\'a}lez, Imad Ez-Zazi, Ahmed El~Oualkadi, and Mounir Arioua.
\newblock A reinforcement learning based transmission parameter selection and energy management for long range internet of things.
\newblock {\em Sensors}, 22(15):5662, 2022.

\bibitem{acosta2023dynamic}
Laura Acosta-Garcia, Juan Aznar-Poveda, Antonio~Javier Garcia-Sanchez, Joan Garcia-Haro, and Thomas Fahringer.
\newblock Dynamic transmission policy for enhancing lora network performance: A deep reinforcement learning approach.
\newblock {\em Internet of Things}, 24:100974, 2023.

\bibitem{sotiriadis2014towards}
Stelios Sotiriadis, Nik Bessis, Eleana Asimakopoulou, and Navonil Mustafee.
\newblock Towards simulating the internet of things.
\newblock In {\em 2014 28th International Conference on Advanced Information Networking and Applications Workshops}, pages 444--448. IEEE, 2014.

\bibitem{bor2016lorasim}
Martin~C Bor and Thiemo Voigt.
\newblock Lorasim.
\newblock [Online]. \url{https://mcbor.github.io/lorasim/} (accessed 08.12.2024), 7 2017.

\bibitem{bor2016lora}
Martin~C Bor, Utz Roedig, Thiemo Voigt, and Juan~M Alonso.
\newblock Do lora low-power wide-area networks scale?
\newblock In {\em Proceedings of the 19th ACM International Conference on Modeling, Analysis and Simulation of Wireless and Mobile Systems}, pages 59--67, 2016.

\bibitem{slabicki2022flora}
Mariusz Slabicki.
\newblock Flora - a framework for lora simulations with omnet++, 2022.
\newblock \url{https://flora.aalto.fi/} [Accessed: 18.05.2024].

\bibitem{slabicki2018adaptive}
Mariusz Slabicki, Gopika Premsankar, and Mario Di~Francesco.
\newblock Adaptive configuration of lora networks for dense iot deployments.
\newblock In {\em NOMS 2018-2018 IEEE/IFIP Network Operations and Management Symposium}, pages 1--9. IEEE, 2018.

\bibitem{magrin2016network}
Davide Magrin.
\newblock Network level performances of a lora system.
\newblock Master's thesis, University of Padua, 2016.

\bibitem{magrin2016ns3}
Davide Magrin, Martina Capuzzo, Stefano Romagnolo, and Michele Luvisotto.
\newblock Lorawan ns-3 module.
\newblock [Online]. \url{https://github.com/signetlabdei/lorawan} (accessed 08.12.2024), 10 2024.

\bibitem{UniCTARS44}
Unict-arslab/lwn-simulator: A lorawan nodes' and network simulator that works with a real lorawan environment (such as chirpstack) and equipped with a web interface for real-time interaction.
\newblock [Online; accessed 2024-12-30].

\bibitem{ta2019lora}
Duc-Tuyen Ta, Kinda Khawam, Samer Lahoud, C{\'e}dric Adjih, and Steven Martin.
\newblock Lora-mab: Toward an intelligent resource allocation approach for lorawan.
\newblock In {\em 2019 IEEE global communications conference (GLOBECOM)}, pages 1--6. IEEE, 2019.

\bibitem{serati2022adr}
Reza Serati, Benyamin Teymuri, Nikolaos~Athanasios Anagnostopoulos, and Mehdi Rasti.
\newblock Adr-lite: A low-complexity adaptive data rate scheme for the lora network.
\newblock In {\em 2022 18th International Conference on Wireless and Mobile Computing, Networking and Communications (WiMob)}, pages 296--301. IEEE, 2022.

\bibitem{marais2019review}
Jaco~M Marais, Adnan~M Abu-Mahfouz, and Gerhard~P Hancke.
\newblock A review of lorawan simulators: Design requirements and limitations.
\newblock In {\em 2019 international multidisciplinary information technology and engineering conference (IMITEC)}, pages 1--6. IEEE, 2019.

\bibitem{acosta2024proactive}
Laura Acosta-Garcia, Juan Aznar-Poveda, Antonio-Javier Garcia-Sanchez, Joan Garcia-Haro, and Thomas Fahringer.
\newblock Proactive adaptation of data rate in mobile lora-based iot devices using machine learning.
\newblock In {\em 2024 IEEE 99th Vehicular Technology Conference (VTC2024-Spring)}, pages 1--5. IEEE, 2024.

\bibitem{acosta2025adrl}
Laura Acosta-Garcia, Juan Aznar-Poveda, Antonio~Javier Garcia-Sanchez, Jakob Hollenstein, Joan Garcia-Haro, and Thomas Fahringer.
\newblock Adrl: A reconfigurable energy-efficient transmission policy for mobile lora devices based on reinforcement learning.
\newblock {\em Internet of Things}, page 101577, 2025.

\bibitem{farhad2022hadr}
Arshad Farhad and Jae-Young Pyun.
\newblock Hadr: A hybrid adaptive data rate in lorawan for internet of things.
\newblock {\em ICT Express}, 8(2):283--289, 2022.

\bibitem{francisco2021improving}
S{\'e}rgio Francisco, Pedro Pinho, and Miguel Lu{\'\i}s.
\newblock Improving lora network simulator for a more realistic approach on lorawan.
\newblock In {\em 2021 Telecoms Conference (ConfTELE)}, pages 1--6. IEEE, 2021.

\bibitem{pop2017does}
Alexandru-Ioan Pop, Usman Raza, Parag Kulkarni, and Mahesh Sooriyabandara.
\newblock Does bidirectional traffic do more harm than good in lorawan based lpwa networks?
\newblock In {\em GLOBECOM 2017-2017 IEEE Global Communications Conference}, pages 1--6. IEEE, 2017.

\bibitem{opensim2019omnetpp}
OpenSim Ltd.
\newblock Omnet++ discrete event simulator.
\newblock [Online]. \url{https://omnetpp.org/} (accessed 18.05.2024), 2019.

\bibitem{opensim2023inet}
OpenSim Ltd.
\newblock Inet framework.
\newblock [Online]. \url{https://inet.omnetpp.org/} (accessed 18.05.2024), 2023.

\bibitem{nsnam2024ns3}
nsnam.
\newblock ns-3 network simulator.
\newblock [Online]. \url{https://www.nsnam.org/} (accessed 08.12.2024), 2024.

\bibitem{centenaro2017impact}
Marco Centenaro, Lorenzo Vangelista, and Ryuji Kohno.
\newblock On the impact of downlink feedback on lora performance.
\newblock In {\em 2017 IEEE 28th Annual International Symposium on Personal, Indoor, and Mobile Radio Communications (PIMRC)}, pages 1--6. ieee, 2017.

\bibitem{callebaut2019cross}
Gilles Callebaut, Geoffrey Ottoy, and Liesbet Van~der Perre.
\newblock Cross-layer framework and optimization for efficient use of the energy budget of iot nodes.
\newblock In {\em 2019 IEEE Wireless Communications and Networking Conference (WCNC)}, pages 1--6. IEEE, 2019.

\bibitem{maartenw68}
maartenweyn/lpwansimulation.
\newblock [Online; accessed 2025-01-09].

\bibitem{eigner2021interference}
Harald Eigner.
\newblock Interference analysis of lorawan systems.
\newblock Master's thesis, TU Wien, 2021.

\bibitem{zhang2023multiagent}
Xu~Zhang, Ziqi Lin, Shimin Gong, Bo~Gu, and Dusit Niyato.
\newblock Multiagent reinforcement learning with an attention mechanism for improving energy efficiency in lora networks.
\newblock In {\em GLOBECOM 2023-2023 IEEE Global Communications Conference}, pages 4152--4157. IEEE, 2023.

\bibitem{yu2024resolve}
Fu~Yu, Xiaolong Zheng, Yuhao Ma, Liang Liu, and Huadong Ma.
\newblock Resolve cross-channel interference for lora.
\newblock In {\em 2024 IEEE 44th International Conference on Distributed Computing Systems (ICDCS)}, pages 1027--1038. IEEE, 2024.

\bibitem{loraalliance2022regionalparameters}
LoRa Alliance.
\newblock Rp002-1.0.4 regional parameters.
\newblock [Online]. \url{https://resources.lora-alliance.org/technical-specifications/rp002-1-0-4-regional-parameters} (accessed 08.03.2024), 2022.

\bibitem{semtech2020sxdatasheet}
Semtech.
\newblock Sx1276/77/78/79 datasheet, rev. 7, 2020.
\newblock \url{https://www.semtech.com/products/wireless-rf/lora-connect/sx1276#documentation} [Accessed: 20.05.2024].

\bibitem{haxhibeqiri2018survey}
Jetmir Haxhibeqiri, Eli De~Poorter, Ingrid Moerman, and Jeroen Hoebeke.
\newblock A survey of lorawan for iot: From technology to application.
\newblock {\em Sensors}, 18(11):3995, 2018.

\bibitem{LoRaSim}
Lorasim.
\newblock \url{https://www.lancaster.ac.uk/scc/sites/lora/lorasim.html}.
\newblock (Accessed on 12/20/2023).

\bibitem{magrin2017performance}
Davide Magrin, Marco Centenaro, and Lorenzo Vangelista.
\newblock Performance evaluation of lora networks in a smart city scenario.
\newblock In {\em 2017 IEEE International Conference on communications (ICC)}, pages 1--7. ieee, 2017.

\bibitem{coutaud2021lora}
Ulysse Coutaud, Martin Heusse, and Bernard Tourancheau.
\newblock Lora channel characterization for flexible and high reliability adaptive data rate in multiple gateways networks.
\newblock {\em Computers}, 10(4):44, 2021.

\bibitem{gao2019towards}
Weifeng Gao, Wan Du, Zhiwei Zhao, Geyong Min, and Mukesh Singhal.
\newblock Towards energy-fairness in lora networks.
\newblock In {\em 2019 IEEE 39th International Conference on Distributed Computing Systems (ICDCS)}, pages 788--798. IEEE, 2019.

\bibitem{amichi2020joint}
Licia Amichi, Megumi Kaneko, Ellen~Hidemi Fukuda, Nancy El~Rachkidy, and Alexandre Guitton.
\newblock Joint allocation strategies of power and spreading factors with imperfect orthogonality in lora networks.
\newblock {\em IEEE Transactions on Communications}, 68(6):3750--3765, 2020.

\bibitem{toro2021modeling}
Ver{\'o}nica Toro-Betancur, Gopika Premsankar, Mariusz Slabicki, and Mario Di~Francesco.
\newblock Modeling communication reliability in lora networks with device-level accuracy.
\newblock In {\em IEEE INFOCOM 2021-IEEE Conference on Computer Communications}, pages 1--10. IEEE, 2021.

\bibitem{xu2022x}
Zhuqing Xu, Junzhou Luo, Zhimeng Yin, Shuai Wang, Ciyuan Chen, Jingkai Lin, Runqun Xiong, and Tian He.
\newblock X-mac: Achieving high scalability via imperfect-orthogonality aware scheduling in lpwan.
\newblock In {\em 2022 IEEE 30th International Conference on Network Protocols (ICNP)}, pages 1--11. IEEE, 2022.

\bibitem{semtech2016adr}
Semtech.
\newblock Lorawan - simple rate adaptation recommended algorithm.
\newblock [Online]. \url{https://www.thethingsnetwork.org/forum/uploads/default/original/2X/7/7480e044aa93a54a910dab8ef0adfb5f515d14a1.pdf} (accessed 02.12.2024), 08 2016.

\bibitem{semtech2023adrbackoff}
Semtech.
\newblock Implementing adaptive data rate.
\newblock [Online]. \url{https://learn.semtech.com/mod/book/view.php?id=174&chapterid=162} (accessed 18.11.2024), 2023.

\bibitem{repository_fast-lora}
juanaznarp94/fast-lora: Fast-lora simulation framework for lora networks.
\newblock [Online; accessed 2025-03-11].

\bibitem{GymDocum55}
Gym documentation.
\newblock [Online; accessed 2025-04-08].

\bibitem{RLlibInd4}
Rllib: Industry-grade, scalable reinforcement learning — ray 2.44.1.
\newblock [Online; accessed 2025-04-08].

\bibitem{PettingZ20}
Pettingzoo documentation.
\newblock [Online; accessed 2025-04-08].

\bibitem{StableBa63}
Stable-baselines3 docs - reliable reinforcement learning implementations — stable baselines3 2.6.1a0 documentation.
\newblock [Online; accessed 2025-04-08].

\bibitem{liando2019known}
Jansen~C Liando, Amalinda Gamage, Agustinus~W Tengourtius, and Mo~Li.
\newblock Known and unknown facts of lora: Experiences from a large-scale measurement study.
\newblock {\em ACM Transactions on Sensor Networks (TOSN)}, 15(2):1--35, 2019.

\bibitem{semtech2017sxdatasheet}
Semtech.
\newblock Sx1272, 2019.
\newblock [Online]. \url{https://www.semtech.com/products/wireless-rf/lora-connect/sx1272\#documentation} (accessed 20.05.2024).

\end{thebibliography}

\end{document}